\documentclass[12pt]{article}
\usepackage{latexsym}
\usepackage{epsfig,amssymb,euscript}
\usepackage{amsmath}
\usepackage{array,calc,epsfig}

\def\be{\begin{equation}}
\def\ee{\end{equation}}
\def\bea{\begin{eqnarray}}
\def\eea{\end{eqnarray}}

\begin{document}
\baselineskip=15.5pt
\pagestyle{plain}
\setcounter{page}{1}
\begin{titlepage}

\leftline{\tt hep-th/0606059}

\vskip -.8cm

\rightline{\small{\tt LPTHE-06-03}}
\rightline{\small{\tt UB-ECM-PF 06/15}}

\begin{center}

\vskip 1.4 cm

{\LARGE{\bf New predictions on meson decays from string splitting}}
\vskip .3cm 

\vskip 1.2cm
\vspace{20pt}
{\large 
F. Bigazzi $^{a}$, A. L. Cotrone $^{b}$}\\
\vskip 1.2cm

\textit{$^a$  LPTHE, Universit\'es Paris VI and VII, 4 place Jussieu; 75005, Paris, France and \\ INFN, Piazza dei Caprettari, 70; I-00186  Roma, Italy.}\\
\textit{$^b$ Departament ECM, Facultat de F\'isica, Universitat de Barcelona and \\ Institut
de Fisica d'Altes Energies, Diagonal 647, E-08028 Barcelona, Spain.}\\
\end{center}

\vspace{12pt}

\begin{center}
\textbf{Abstract}
\end{center}

\vspace{4pt}{\small \noindent We study certain exclusive decays of high spin mesons into mesons in models of large $N_c$ Yang-Mills with few flavors at strong coupling using string theory. The rate of the process is calculated by studying the splitting of a macroscopic string on the relevant dual gravity backgrounds. In the leading channel for the decay of heavy quarkonium into two open-heavy quark states, one of the two produced mesons has much larger spin than the other. 
In this channel the decay rate is only power-like suppressed with the mass of the produced quark-anti quark pair. We also reconsider decays of high spin mesons made up of light quarks, confirming the linear dependence of the rate on the mass of the decaying meson. As a bonus of our computation, we provide a formula for the splitting rate of a macroscopic string lying on a Dp-brane in flat space. 
}

\vfill \vskip 5.mm \hrule width 5.cm \vskip 2.mm {\small \noindent e-mail: bigazzi@lpthe.jussieu.fr, cotrone@ecm.ub.es}

\noindent
\end{titlepage}

\newpage


\section{Introduction}
\label{intro}
Many efforts are being recently devoted to build a bridge between string theory predictions and QCD phenomenology. Though we do not have yet a string model describing real QCD, we can try and understand how far we can go with the available setups. 
These
become manageable when the number of colors $N_c$ is formally taken to infinity instead of taking it equal to three. Other relevant sources of discrepancy with real QCD are model dependent, but in general are related to the fact that, in the string/gravity descriptions we are able to manage, 
the interesting gauge theory sector is always coupled to spurious fields (like the Kaluza-Klein modes coming from extra dimensions) and asymptotic freedom is not properly realized (though the gauge coupling can grow at low energies, the UV 't Hooft coupling $\lambda$ has to be taken very large in order for the geometries to have small curvatures). Nevertheless it is important to put our models at work, both with the objective of quantitatively determine how far we are from reality and to produce, in the limit of our analysis, falsifiable predictions.

Hadronic physics is a natural arena to critically study the relevance of string theory in accounting for the strong coupling dynamics of QCD. In this letter we focus on 
high spin mesons decaying into mesons in large $N_c$ models of strongly coupled Yang-Mills, with $N_f$ flavors in the regime where $N_f$ is taken to be much smaller than $N_c$. This is a realm where analytical predictions on certain exclusive decay rates can be extracted from string theory. Unfortunately, at the moment, it is also a regime which lies far from the available experimental data in real QCD. Nevertheless a critical comparison of the results for the string dual models with phenomenological effective theories or lattice predictions for QCD in the future is not excluded in principle.
Let us recall again that the string dual models are reliable in the planar limit at strong coupling and are generically coupled to extra fields beyond the YM and quark ones.
As such, they are far from being exact descriptions of QCD.
Nevertheless, they have revealed to encode many properties of QCD (see for example \cite{myers,ss}) and dynamical process in these theories in principle could share some qualitative behavior with real world QCD.
  
In the first part of the paper we focus on high spin heavy quarkonia\footnote{Our results are trivially extended to high spin mesons made up of two different heavy quarks.} which lie above the threshold for open-heavy quark production. The rate for a decay of the kind $\bar Q Q \rightarrow \bar Q q + Q\bar q$ is evaluated  by studying the splitting of a macroscopic semiclassical string, attached to a ``Q-flavor'' D-brane, on a  ``q-flavor'' D-brane which the string intersects. To be concrete, we will mostly adapt the general results obtained in \cite{noi} to a precise string model for a quenched Yang-Mills theory \cite{myers} with $N_c$ D4 branes wrapped on a cycle and $N_f$ D6 ``flavor'' branes. Some of the necessary ingredients to follow our calculations, a brief discussion on the features of the string models we are going to use and the regimes of reliability of our analysis will be given in Section 2. 

As we are going to show in Section 3, 
in the regime we consider, the leading decay channel for heavy quarkonia
at strong coupling is asymmetric i.e. one of the produced mesons has a much higher spin than the other (this is possibly a genuine strong coupling effect).  
The decay rate at leading order has only a power-like suppression with the mass $m_q$ of the produced quark-antiquark pair. 
In contrast, symmetric decays, characterized by the fact that the two outgoing mesons have spin of the same order, are exponentially suppressed with the (squared) mass $m_q$.
In this sense, the latter channel, though phenomenologically implemented within the Lund model of hadron fragmentation \cite{Lund}, and studied in the string setup in \cite{zamaklar}, is subdominant w.r.t. the channel considered in this letter\footnote{The rate for the decay via emission of massless modes in flat space has been calculated in \cite{iengo} to be $\Gamma\sim g_s M^{5-D}$, where $D$ is the number of non compact dimensions and $M$ the mass of the state.} at least in the strongly coupled regime $\lambda\gg1$. 

Another relevant feature of the decay rate we calculate is that it depends on the spin of the decaying meson only in the laboratory reference frame, where it is suppressed by the spin itself. 
All these properties (though probably not the precise power of $m_q$) should be model independent, i.e. should be common to every stringy description of quenched planar Yang-Mills including flavors\footnote{ 
It would be interesting to test this statement performing a similar calculation in other models that encode some QCD physics \cite{erdmenger}.}.
This is relevant since our results can be possibly falsified. 

It would be interesting if decays of the kind we are about could be studied in the same theories by a WKB treatment of  non relativistic models for heavy quarkonia (see \cite{brambilla} for a state of the art review on the physics of these states), like the Cornell Coupled Channel ($C^3$) one \cite{cornell}. In this model the heavy quarks are treated as non relativistic particles interacting essentially via a linear potential; moreover the model takes into account the possible mixing and interactions between ${\bar Q}Q$ and ${\bar Q}q + {\bar q}Q$ states. 
Predictions on the spectrum of the quarkonia as well as on the exclusive decay rates for states above the threshold have been made (provided certain parameters are fitted with real data), at least for small spin states (see \cite{cornell2} for recent analysis). It should be interesting to put the $C^3$ model at work in the very high spin case.

In a second part of the paper (Section 4) we revisit the decay $\bar q q \rightarrow \bar q q + q\bar q$ of high spin mesons made of light quarks in the string theory setup which seems more adequate, 
namely the one with $N_c$ D4 and $N_f$ D8-branes \cite{ss}. In this case we confirm that the rate is linear with the mass of the decaying meson, as in the old string models. The difference w.r.t. the latter calculations is in a damping factor, that is model-dependent.
As a by-product of our investigation, we provide the splitting rate of a macroscopic open string living on a generic Dp-brane in flat space.

We conclude with a summary and a discussion of the results in Section 5.

\section{Background material: mesons in string theory}

We work in the two known regular string models dual to strong coupling planar quenched QCD coupled to adjoint matter \cite{myers}, \cite{ss}.
They are both based on the same supergravity background generated by D4-branes wrapped on a supersymmetry-breaking cycle \cite{WYM}:
\be\label{metric}
ds^2=(\frac uR)^{3/2} (dx_\mu dx^\mu + \frac{4R^3}{9u_h}f(u)d\theta_2^2) + (\frac{R}{u})^{3/2}  \frac{du^2}{f(u)} + R^{3/2}u^{1/2} d\Omega_4^2,\quad e^{\Phi}=g_s\Bigl( \frac{u}{R}\Bigr)^{3/4},
 \ee
with $f(u)=(u^3-u_h^3)/u^3$, where $h$ stands for ``horizon'' and $u\in [u_h,\infty)$. Moreover, there are $N_c$ units of $F_4$ flux through $S^4$. The string parameters can be expressed in terms of field theory quantities with the relations \cite{bcmp}
\be
u_h=\frac{\lambda m_0 \alpha'}{3},\qquad g_s=\frac{\lambda}{3\pi N_c m_0 \sqrt{\alpha'}},\qquad R^3=\frac{\lambda\alpha'}{3m_0},\qquad T=\frac{\lambda m_0^2}{6\pi}, 
\ee
where $m_0$ is the glueball and adjoint KK matter scale (these cannot be separated if we require that the above background has small curvature), $\lambda=g^2_{YM}N_c$ is the 't Hooft coupling at the UV cut-off 
and $T$ is the effective string tension.
Note that, as opposed to pure Yang-Mills, there are two distinct mass scales, $m_0$ and $\sqrt{T}$.
If $\lambda \gg 1$ the background (\ref{metric}) faithfully describes pure large $N_c$ Yang-Mills at low energy, coupled to adjoint KK matter.

In the quenched approximation, flavors are introduced as $N_f$ ``probe D-branes'', thus not affecting the metric above.
The mesons in this picture are just excitations of the $N_f$ D-branes.

In the model considered in \cite{myers}, the probes are $N_f$ D6 branes embedded in the geometry (\ref{metric}), extended in the $u$ radial direction from a certain value $u_Q$ up to infinity. The flavor symmetry here is $U(N_f)$ by construction. The spontaneous breaking of the chiral $U(1)_A$ symmetry is accounted for by the bending of the flavor branes.
The ``constituent mass'' of the associated quarks is related to $u_Q$. It turns out that $u_Q$ cannot be the extreme value $u=u_h$, but there exists a minimal radial position $u_{min} > u_h$. 
Even if the bare mass of the quarks is zero, the fact that $u_{min} > u_h$ produces an effective non-vanishing constituent mass $m_Q$. 
This is identified with the the energy of an hypothetical string  stretching along the radial coordinate from $u=u_Q$ to the horizon at $u_h$\footnote{As a curiosity we notice that the running coupling of the theory is \cite{bcmp} $g_{YM}^2(u) = (\lambda/N_c)[1-(u_h/u)^3]^{-1/2}$, this inducing a RG-like relation $m'(u)=(N_c/2\pi\alpha'\lambda) g_{YM}^2(u)$.}
\be
m(u)= {T\over m_0}\int_{1}^{u/u_h} dz \left[1-{1\over z^3}\right]^{-{1\over2}}.
\label{massq}
\ee 

In a second model, proposed in \cite{ss}, flavors are dual to $N_f$ curved D8-branes orthogonal to the circle $S^1(\theta_2)$ and extending up to the horizon $u=u_h$; at large $u$ the ``extrema'' of each curved D8-brane is seen as a brane-antibrane pair, thus realizing a dynamical UV restoration of the $U(N_f)\times U(N_f)$ chiral symmetry; this model describes effectively massless quarks.

While small fluctuations of the probe branes describe low-spin mesons in both models, heavy mesons with very high spin ($J\gg\lambda$) in the first scenario \cite{myers} are described by almost $U$-shaped macroscopic strings \cite{crucco} attached to the branes at $u=u_Q$, extending all the way up to a minimum $u=u_0$ and spinning in the Minkowski directions, see figure 1.
\begin{figure}
\begin{center}
\scalebox{0.7}{\includegraphics{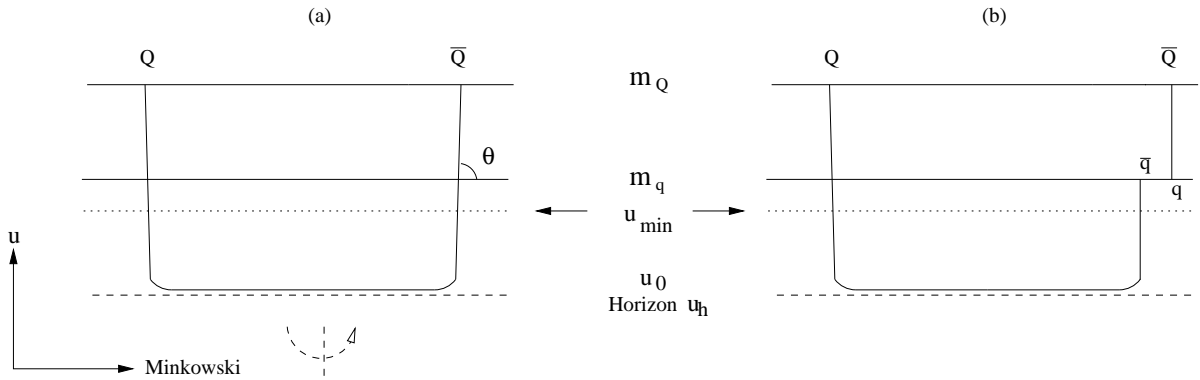}}
\caption{\small{(a) A large spin meson, bound state of two quarks of large mass $m_Q$, described by a string with both end-points on the same brane. The string reaches a minimal position $u_0$, and intersects a second brane, corresponding to lighter quark masses $m_q$. (b) The strings after the splitting, representing two meson bound states of a heavy quark and a light quark.}}
\label{figura1}
\end{center}
\end{figure}
In the second model \cite{ss}, they are just macroscopic open strings extending and spinning in Minkowski (thus lying on the flavor branes) giving rise to linear Regge trajectories.
Crucially, in both cases, in the regime $J\gg \lambda$ the string can be studied semi-classically, the quantum corrections being suppressed.   

The splitting of this macroscopic string is thus interpreted as the exclusive decay of mesons $Q\bar Q \rightarrow Q\bar q + q\bar Q$.
When the quarks $Q$ are heavy we can study the phenomenon with the D6-brane embedding \cite{myers}.
In order for the string to split and the decay to happen, there must be a second type of flavor D6 brane whose minimum is at a lower position $u_q<u_Q$.
Then the string intersects this brane and can split into two strings stretching between the two branes, representing the two decay product mesons. The bending string corresponds to a high spin ($J_1\gg\lambda$) meson, while the other one is 
related to a meson with much smaller spin ($1\ll J_2\ll \lambda$), see \cite{pt}. Note that 
 the splitting can happen only at one of the two intersection points. 

When the quarks are light, we can use the D8 embedding \cite{ss}.
Since the string lies on the brane, it can split at any point. 

We perform a calculation of the decay rates for both splitting processes in string theory.
The strategy \cite{jjp,noi} is to first compute the rates on a torus.
Since the curvature of the background is small and the splitting process is local, the flat space results are good approximations of the actual rates. 

\section{Decay of high spin, very massive quark mesons}

In this section we estimate the decay rate for the process $Q\bar Q \rightarrow Q\bar q + q\bar Q$ of heavy quark high spin mesons in the D6 model \cite{myers}.
The rate on flat space for the splitting of a string intersecting a Dp-brane at generic angle $\theta$, was computed in \cite{noi} to be
\be\label{rate}
\Gamma = \frac{g_s}{16\pi\sqrt{\alpha'}}\frac{\cos^2{\theta}}{\sin{\theta}}\frac{(2\pi\sqrt{\alpha'})^{8-p}}{V_{\perp}},
\ee
where $V_{\perp}$ is the volume of the transverse space.
It can be used on a weakly curved background with (see figure 1)
\be\label{angolo}
\cos^2{\theta}=\frac{r'(u_q)^2}{e^{B-A}+r'(u_q)^2},
\ee
$u_q$ being the radial position of the light-flavor brane $q$, for a string embedding $r(u)$ and a metric in the form
\be
ds^2=e^A (-dt^2 + dr^2 + r^2 d\varphi^2 + dx_3^2) + e^B du^2+...
\ee
In order to interpret the result (\ref{rate}) in field theory language, we apply it to the model at hand.
First of all, since we are on a curved space, we must take the corrected string tension and dilaton, that is $\alpha'\rightarrow \alpha'_{eff}=\alpha'/$warp factor$=\alpha' (\frac{R}{u_q})^{3/2}$, and $g_s\rightarrow e^{\Phi}(u_q)=g_s(\frac{u_q}{R})^{3/4}$.
Second, in this model $p=6$ and the transverse volume can be estimated to be\footnote{In this case, as in \cite{noi}, it is not possible to give a better estimate of the effective volume with a local quantum computation along the lines of \cite{jjp}.}
\be
V_{\perp}=2\pi R_{\theta_2} \cdot 2\pi R_{S^4}= \frac{8\pi^2u_q}{3u_h^{1/2}}R^{3/2}f^{1/2}(u_q).
\ee
Next we evaluate the angular part using, as in \cite{noi}, the expression for the static Wilson loop almost-$U$-shaped profile for the string (see the expression, for example, in \cite{pt})
\be
r'_{st}(u)= (Ru_0)^{3/2}\frac{1}{\sqrt{(u^3-u_0^3)(u^3-u_h^3)}},
\ee
where $u_0$ is the minimal radial position reached by the string.
This choice is justified by the fact that the actual profile of a high spin meson is very well approximated by such Wilson line \cite{crucco,pt}. 
It also means, on the other hand, that we are effectively pushing the mass $m_Q$ of the heavy quarks to infinity.
Putting it in the formula (\ref{angolo}) one gets
\be
\cos^2{\theta}=(\frac{u_0}{u_q})^{3}, \qquad\qquad \frac{\cos^2{\theta}}{\sin{\theta}}=\frac{u_0^3}{u_q^3\sqrt{1-\frac{u_0^3}{u_q^3}}}.
\ee
The rate is suppressed because the angle $\theta$ is equal to $\pi/2$ up to small corrections \cite{noi}.
Here we see that the corrections are now estimated to be power-like with $u_q$.

Putting everything in (\ref{rate}) and multiplying it by a factor $2$ in order to account for the two possible points where the splitting can happen, one gets the following holographic expression for the $\bar Q Q\rightarrow \bar Q q + \bar q Q$ decay rate
\be\label{intermedio}
\Gamma = \frac{\lambda}{16\pi^2 N_c m_0 R^3}\frac{u_h^{1/2}u_0^3 u_q^{1/2}}{\sqrt{(u_q^3-u_0^3)(u_q^3-u_h^3)}}.
\ee
Just note that 
for $u_q\rightarrow \infty$ the rate vanishes as it should.

Up to now, the only approximation made was to use the profile of the string given by the Wilson loop, that is, to consider $u_Q\gg u_h$ and so (see eq. (\ref{massq})) $m_Q \gg T/m_0$. 
Now, in order to write down the rate (\ref{intermedio}) only in terms of field theory quantities, we are going to analyze some definite range of parameters, 
performing some approximations.

The quantity
$u_0$ encodes informations on the spin $J$  
of the quarks forming the high spin meson. 
In the semi-classical regime we work in, $u_0$ is equal to $u_h$ up to exponential terms, namely  \cite{sonne}
\be\label{u0uh}
u_0\sim 
u_h(1+e^{-\frac{3m_0L}{2}}) ,
\ee 
where $L$ is the distance between the two quarks.
This is valid in the regime where the semi-classical approximation is good, that is large $L$.
Now, if we make finite the mass of the quarks, we can express $L$ in terms of $m_Q$ and $J$  \cite{crucco}.
There are two extremal regimes.
One is when the constituent mass of the quarks is very high w.r.t. the energy stored in the flux tube, $m_Q \gg TL$, case in which the mass of the meson is $M\sim 2m_Q + TL \sim 2m_Q$ and the spin is $J\sim 2\sqrt{Tm_Q}L^{3/2}$.
Note that the flux tube has very small energy and the energy of the meson is ``concentrated'' at its end-points, near the quarks.
The limit $m_Q \gg TL$ means $m_Q^2 \gg TJ$. 
Nevertheless, in order to have a long string $m_0L \gg 1 $ (here we measure it w.r.t. the mass scale $m_0$), it must be that $m_0 J^{2/3} \gg (Tm_Q)^{1/3}$, that is, taking into account $m_Q \gg TL$, $J \gg T/m_0^2 \sim \lambda$. 
This is just the high spin request $J \gg \lambda$.

The opposite limit is when the quarks are mildly massive, i.e. they are light w.r.t. the flux tube ({\emph{not}} w.r.t. $\Lambda_{QCD}$), $m_Q \ll TL$, for which $M \sim TL$ and $J\sim T L^2$.
In this case too $m_0L$ is very large for $J \gg \lambda$.
Note that in this case the energy of the meson is mainly stored in the flux-tube. 
Even if the mass $m_Q$ is very large ($m_Q\gg T/m_0$), taking $J$ sufficiently large one can easily reach this limit. 
Nevertheless, in the string picture, even if the flux tube is very long and stores practically all the energy of the meson, its energy density is still smaller than the one allowing for pair-production, so, semiclassically, the splitting can happen again only at the extrema of the tube, 
where the energy density rapidly increases
\cite{noi}.

Thus in all the cases,  
in the regime where $J \gg \lambda$, 
up to exponential factors, $u_0=u_h$.
In making this approximation we loose
some informations on $J$.
This is to be expected: in this semi-classical regime the string has always the ``straight'' shape and always reaches the horizon, irrespective of the details of the meson.

The rate (\ref{intermedio}) is thus, up to exponential terms (that can always be included multiplying by the appropriate factor (\ref{u0uh})),
\be
\Gamma = \frac{\lambda}{16\pi^2 N_c m_0 R^3}\frac{u_h^{7/2} u_q^{1/2}}{u_q^3-u_h^3}={\lambda m_0\over 16\pi^2N_c}{\sqrt{x}\over(x^3-1)},
\ee
where we have introduced the dimensionless variable $x=u_q/u_h$.

Up to now we have pushed $u_Q$ to infinity, so the resulting rate was independent on $u_Q$. Since this variable is ultimately related with the mass of the constituent quarks of the original meson, it is important to refine our calculation to include the leading order effect due to the big, but finite, quark mass. In the limit $y\equiv(u_Q/u_h)\gg 1$, the actual profile of a high spin meson is a small perturbation of that of a static Wilson line: $r'(u)=r'_{st}(u)+\delta r'(u)$ (see \cite{pt}) with
\be
\delta r'(u) \sim \frac{u^3}{(u^3-u_h^3)^2}[-R^3\omega^2 \frac{r_{st}}{1-\omega^2 r_{st}^2}(u-u_h)],
\ee
where $r_{st}=L/2$ (this is true for almost every $u$ apart from those values exponentially close to $u_h$) and
\be\label{angvel} 
\omega^2 = \frac{T}{(L/2) \tilde M_Q + (L/2)^2 T}\, , \qquad \tilde M_Q = \frac{1}{2\pi\alpha'}\frac{(u_Q-u_h)u_Q^3}{u_Q^3-u_h^3}\sim \frac{1}{2\pi\alpha'}u_Q\, .
\ee

All in all we can write
\be
r'(u)\approx {(Ru_h)^{3/2}\over u_h^3(x^3-1)}\left[1-{x^3(x-1)\over y(x^3-1)}\right],
\ee
where the second term in parenthesis accounts for the leading order correction of the profile in terms of the mass of the meson. We can now refine our evaluation of the angular part of the decay rate by using formula (\ref{angolo}). The final result is
\be
\label{ratef}
\Gamma={\lambda m_0\over 16\pi^2N_c}{\sqrt{x}\over(x^3-1)}\left[1+ {1\over y}{(x-1)(1-2x^3)\over(x^3-1)}\right].
\ee 
In order to give a clear interpretation of this formula, we will express it in terms of the quark masses (\ref{massq}), focusing on two limits where analytical expressions can be given. 

The first is when 
$u\gg u_h$ (large $x$), where $m_q\approx u_q/(2\pi\alpha')\gg T/m_0$. 
The resulting rate, expressed in terms of the quark masses, reads
\be
\label{finalrate1}
\Gamma \sim \frac{\lambda}{16\pi^2 N_c} \left(\frac{T}{m_0}\right)^{5/2} \frac{m_0}{m_q^{5/2}}\left[1-2{m_q\over m_Q}\right].
\ee
Of course, since there are two scales, the string tension $T$ and the squared glueball mass $m_0^2$, the rate can be rewritten in a number of equivalent ways, playing at will with them and the coupling\footnote{Just note that we could formally try and extrapolate to a theory closer to real QCD taking $T\sim m_0^2\sim\Lambda_{QCD}^2$. The rate would go as
$
\Gamma \sim\frac{\lambda}{N_c}\frac{\Lambda_{QCD}^{7/2}}{m_q^{5/2}}\left[1-2{m_q\over m_Q}\right].
$ We will refer to this formal limit as the ``QCD limit''.}.

The opposite limit is when $x\approx 1$ i.e. when $u_q$ is sufficiently close to $u_h$ (but not so close to cross the minimal distance that the flavor branes are allowed to reach\footnote{The minimum value for $u_q/u_h$ is around $1.04$.}). In this case at leading order we get
\begin{equation}
m_q\approx \left({T\over m_0}\right){2\over\sqrt{3}}\sqrt{{u-u_h}\over u_h},
\end{equation}
as can be easily deduced from the energy of a string stretching in the near horizon limit of the metric (see \cite{bcmp} for its expression). Thus in this case the decay rate goes as
\begin{equation}
\label{finalrate2}
\Gamma \sim {\lambda\over 36\pi^2N_c}\left({T\over m_0}\right)^2 {m_0\over m_q^2}\left[1-{T\over 3m_0 m_Q}\right],
\end{equation}
with a different dependence on $m_q$ w.r.t. the other limit\footnote{In the ``QCD limit'' $\Gamma \sim {\lambda\over N_c} {\Lambda^3_{QCD}\over m_q^2}\left[1-{\Lambda_{QCD}\over m_Q}\right]$.}.

We report in figure 2 the interpolation of the general formula (\ref{ratef}) between the two behaviors (\ref{finalrate1}), (\ref{finalrate2}) (corresponding to large and small $m\equiv m_qm_0/T$ respectively) for the ratios of masses $m_q/m_Q$ corresponding to that of the quarks $u,s,c$ with the $c,b$ ones.

\begin{figure}
\begin{center}
\scalebox{0.7}{\includegraphics{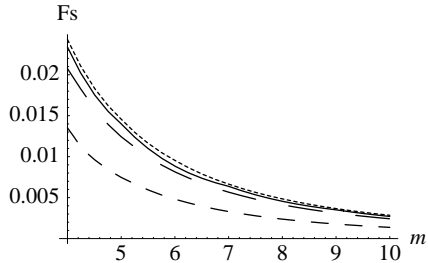}}
\caption{\small{The functions ${\rm Fs}\equiv {\sqrt{x}\over(x^3-1)}\left[1+ {1\over y}{(x-1)(1-2x^3)\over(x^3-1)}\right]$ plotted against $m\equiv m_qm_0/T$ for the ratios of quark masses (from the top to the bottom): $m_u/m_b\ $ (and $m_u/m_c$),$\  m_s/m_b,\  m_s/m_c,\  m_c/m_b$.}}
\label{figura2}
\end{center}
\end{figure}

In the laboratory reference frame these rates must be multiplied by the relativistic $\sqrt{1-v^2}$ factor.
Since the decay can happen only around the heavy quarks one can approximate with $L/2$ the distance of the splitting point from the center of rotation, so that $\sqrt{1-v^2}=\sqrt{1-(\omega L/2)^2}$.
Then, using formula (\ref{angvel}) this factor reads \cite{talk}
\be\label{relativ1}
\sqrt{\frac{2m_Q/L}{T+2m_Q/L}}.
\ee
Remember that $L$ is proportional to some power of $J$.
In the regime where the decaying meson mass is mainly due to the quark mass, $2m_Q \gg LT$, the factor above gives just a subleading correction to the rate.
In this regime, as we already recalled, $L=J^{2/3}/(4m_Q T)^{1/3}$, so that the rate reads
\be\label{relativ2}
\Gamma \bigl(1-\frac{(TJ)^{2/3}}{(4m_Q)^{4/3}}  \bigr).
\ee
In the opposite regime $2m_Q \ll LT$, where the flux tube is much heavier, instead, the relativistic factor introduces a relevant dependence of the rate on $m_Q$ and $J$.
In fact, using the relation for the leading Regge trajectory $L=\sqrt{8J/\pi T}$, the rate is
\be\label{relativ3}
\Gamma \sqrt{m_Q\sqrt{\frac{\pi}{2T}}}\frac{1}{J^{1/4}}.
\ee
It is suppressed by the large spin $J$. 

Let us comment on the formulas for the decay.
The first thing to notice, for every value of $x$, is that the rate
has the right behavior with $N_c$ (trivially), the right qualitative behavior with $\lambda$ (it grows with $\lambda$) and a inverse behavior with $m_q$: as the latter grows (remember it can never be really zero), the rate decreases.
A second important thing is that
the leading order suppression of the rate with the mass $m_q$ is
only {\it power-like} and not exponential, so this happen to be the {\it leading decay channel} in this model.

The corrections to the rate due to the finite quark masses $m_Q$ vanish for $m_Q\rightarrow \infty$; 
moreover $\Gamma(x\gg1)$ increases as $m_q/m_Q$ decreases. 
In the $x\sim1$ limit, i.e. in the limit of small mass for the produced quark antiquark pair, the suppression is w.r.t. the IR scales $(T/m_0)$.

The spin $J$ enters in the expression for the rate only through the relativistic factor in the $2m_Q \ll LT$ regime, where as said it suppresses the process. 
This is reminiscent of the suppression due to the centrifugal barrier in some phenomenological models \cite{spindep}. 
Instead, the corrections (\ref{u0uh}) are exponentially suppressed with $J$.

As a final remark, note that that the mild dependence on the mass 
$m_Q$
and the fact that for high spin mesons made of different heavy quarks the shape of the string is again ``straight'', imply that the very same result
for the decay of heavy quarkonia applies to more general mesons made up of different heavy quarks.

\section{Decay rate for a string on a Dp-brane and light quark high spin mesons}

In this section we estimate the decay rates for the process $q\bar q \rightarrow q\bar q + q\bar q$ of light quark high spin mesons in the D8 model \cite{ss}.
In order to do so, we first calculate the rate on flat space for the splitting of a string lying on a Dp-brane.
The calculation is very similar to the one in \cite{dp,jjp,noi}, so we will just sketch it and present the small differences, addressing the interested reader to those papers for details.

We consider a string stretched along a dimension $X$ on a Dp-brane on a compact space $R_t \times X \times T_{\|} \times T_{\perp}$ where $X$ has the length of the string $L$ and the rest of the space on the brane (transverse to the brane) $T_{\|}$ ($T_{\perp}$) has volume $V_{\|}$ ($V_{\perp}$).
The amplitude and the vertex operators are just as in \cite{noi}, with volume $V=L  V_{\|} V_{\perp}$.
Also the kinematic terms are almost the same, $p_L^2=p_R^2=2/\alpha'$, $p_{L,R}=p \pm \vec L/2\pi\alpha'$, $\vec L=(0_t,L,\vec 0_{\|}, \vec 0_{\perp})$.
For simplicity, we work in the rest frame of the string, so that $p=m(1,0,\vec 0_{\|}, \vec 0_{\perp})$, $m^2=(L/2\pi\alpha')^2-2/\alpha'$.

The invariants that appear in the amplitude, considering that the open string metric is $G^{\mu\nu}={\rm diag}(-1,1,1_{\|},-1_{\perp})$, are just
\bea
-\sigma &\equiv &\frac{\alpha'}{2}p_L G p_R 
=-\alpha'(-\frac{1}{\alpha'}+(L/2\pi\alpha')^2) \quad \Rightarrow \quad \sigma = -1+ \alpha' (L/2\pi\alpha')^2 \nonumber\\
-1-\frac{\alpha' t}{4}&\equiv & \frac{\alpha'}{2}p_L p'_L 
=-1 \quad \Rightarrow \quad t=0\nonumber\\
\sigma-\frac{\alpha' t}{4}&\equiv & \frac{\alpha'}{2}p_L G p'_R 
=-1+ \alpha' (L/2\pi\alpha')^2
\eea
(note that $p'_{L,R}=-p_{L,R}$).
Keeping $t\neq 0$ as a regulator one finds the very same integral as in \cite{noi}, so that in the Regge limit the imaginary part of the amplitude is finite and reads
\be
{\rm Im} {\cal M} \sim N_{D^2}\frac{k^2\pi \sigma}{(2\pi)^2 V},
\ee
where $N_{D^2}=2 \pi^2L V_{\|}/(2\pi)^p(\alpha')^\frac{p+1}{2} g_s$.
The long string limit $\sigma \sim \alpha' (L/2\pi\alpha')^2$, $m\sim L/2\pi\alpha'$ finally gives
\be\label{rateflat}
\Gamma =\frac{1}{m}  {\rm Im} {\cal M} = \frac{g_s}{8(2\pi)^2 \alpha'} \frac{(2\pi \sqrt{\alpha'})^{9-p}}{V_{\perp}} L.
\ee
This is exactly the expected behavior: since the string is entirely on the brane, it can split at any point, so the rate is proportional to its length $L$. The suppression factor $(2\pi \sqrt{\alpha'})^{9-p}/V_{\perp}$ is due to the quantum delocalization of the string in the directions transverse to the brane.

The formula above can also be extracted\footnote{We are grateful to Luca Martucci for this observation.} as a special interesting limit of formula (\ref{rate}), which is divergent for $\theta\rightarrow0$; this depends on the fact that in this case the torus used for the calculation becomes singular. But if we impose that the direction of length $L\sin\theta$ in the limit becomes one of the transverse directions we can write
$V_{\perp(8-p)}= V_{\perp(9-p)} / L \sin\theta$. By making this substitution in (\ref{rate}) we get the interpolating rate
\be\label{rateinterpol}
\Gamma = \frac{g_s}{32\pi^2 \alpha'} \frac{(2\pi \sqrt{\alpha'})^{9-p}}{V_{\perp(9-p)}} L \cos^2\theta,
\ee
which for $\theta\rightarrow0$ exactly gives (\ref{rateflat}).

\subsection{Decay rate for light quark high spin mesons}

In order to apply formula (\ref{rateflat}) to our D8 case, we first estimate the suppression due to the transverse dimension, that is one-dimensional.
We follow the procedure of \cite{jjp} literally, evaluating the quantum delocalization due to the quadratic fluctuations of the world-sheet massive field associated to that direction.
Near the horizon the $u,\ \theta$ part of the metric (\ref{metric}) is just flat ${\mathbb R}^2$ and the branes fill just one of the two directions, the other one being the one we are after.
Specifically, the metric in this region reads (see formula (2.5) in \cite{bcmp})
\be
ds^2\sim \left( \frac{u_h}{R} \right)^{3/2} [1 + \frac{9 y_iy_i}{8R^{3/2}u_h^{1/2}}] dx_\mu dx^\mu + dy_i dy_i + ...
\ee 
where we defined $\frac43 (R^{3/2}u_h^{1/2})(dr^2 +r^2 d\theta^2)\equiv dy_i dy_i$.
In the notation of \cite{jjp} the potential for the transverse world-sheet mode is thus
\be
\frac{1}{2\pi\alpha'}\left( \frac{u_h}{R} \right)^{3/2}[1 + \frac{9 y_iy_i}{8R^{3/2}u_h^{1/2}}],
\ee
so that the fluctuations create a broadening 
\be
\omega = \log [1+ \frac{4R^{3/2}u_h^{1/2}}{9\alpha'}].
\ee
Note the logarithmic behavior with $R^{3/2}u_h^{1/2}$ as opposed to the linear one in the metric.
Inserting the field theory expressions we conclude that
\be
\frac{(2\pi \sqrt{\alpha'})^{9-p}}{V_{\perp}}= \frac{2\pi}{\log^{1/2}(1+\frac{8\pi T}{9m_0^2})}.
\ee 

As final preliminaries for the translation of formula (\ref{rateflat}), note that
\be
\frac{g_s}{\alpha'} \rightarrow \frac{e^{\Phi}}{\alpha'_{eff}}=\frac{g_s}{\alpha'}\left( \frac{u_h}{R}\right)^{\frac{9}{4}}=\frac{\lambda}{N_c}\frac{m_0^2\lambda^{3/2}}{3^{5/2}\pi} 
\ee
and that for the strings on the leading Regge trajectory
\be
L=\sqrt{\frac{8 J}{\pi T}}=\frac{2 M}{\pi T},
\ee
where $M$ is the meson mass (the energy of the string).

Putting everything in  (\ref{rateflat}) one gets
\be\label{gamma2}
\Gamma= \frac{\lambda}{N_c} \frac{1}{6\pi}\frac{1}{\log^{1/2}(1+\frac{8\pi T}{9m_0^2})}\frac{T}{m_0}\sqrt{J}
\ee
or equivalently
\be\label{finalrate3}
\Gamma= \frac{\lambda}{N_c} \frac{1}{6\pi \sqrt{2\pi}}\frac{1}{\log^{1/2}(1+\frac{8\pi T}{9m_0^2})}\frac{\sqrt{T}}{m_0}M.
\ee
In the laboratory reference frame the rate per unit length $\Gamma/L$ must be multiplied by the relativistic $\sqrt{1-v^2}$ factor and then it must be integrated along the length of the string.
At each point $z$ of the string one has $\sqrt{1-v^2}=\sqrt{1-z^2\omega^2}=\sqrt{1-z^2(2/L)^2}$.
Integrating $\Gamma\sqrt{1-z^2(2/L)^2}/L$ for $z\in [-L/2,L/2]$ gives just $\Gamma\pi/4$ \cite{dp}, thus not altering the qualitative form of the rate (\ref{finalrate3}).

All in all, this confirms the expected behavior of the decay rate with the coupling and $N_c$ (obviously), and, most importantly, with the mass $M$ of the decaying meson\footnote{In the ``QCD limit" this is just $\Gamma=\lambda M / N_c$ up to a number.}.

\section{Summary -- phenomenology}

We considered the two available non-singular string realizations of quenched QCD (coupled to KK adjoint matter) in the large $N_c$ limit and at strong (UV) coupling $\lambda$. 
We analyzed the leading channel for the decay of high spin ($J\gg \lambda$) mesons.

The two realizations allow to explore two different mesons.
For large quark mass mesons (large or small are w.r.t $\Lambda_{QCD}$) the decay is of the form $Q \bar Q \rightarrow Q\bar q + \bar Q q$, where $Q$ is a heavy quark of mass $m_Q$ and $q$ is a quark of mass $m_q$.
For small quark mass mesons the decay is of the form $q \bar q \rightarrow q\bar q + \bar q q$, that is, only into mesons made of quarks of the same small mass.\\ 

The decay rate for the high spin, large quark mass mesons (\ref{ratef}), (\ref{finalrate1}), (\ref{finalrate2}) and figure 2, depends on the spin $J$ of the meson only through the relativistic factor (\ref{relativ1}): it is suppressed by $J$, see (\ref{relativ3}), in the regime where $m_Q \ll M$, $M$ being the meson mass.   
The other corrections that include the dependence on the spin (\ref{u0uh}) are exponentially suppressed with the spin itself. 

The rate dependence on the mass $M$ of the meson is mild in the regime where the flux tube is lighter than the quarks, i.e. when $M \sim 2m_Q$.
Then 
at leading order 
the dependence is just
by the factor $(1-2m_q/m_Q- (TJ)^{2/3}/(4m_Q)^{4/3})$ depending on the mass of the lighter quarks $m_q < m_Q$ when $m_q$ is large (\ref{finalrate1}), or by the factor $(1-T/3m_0m_Q- (TJ)^{2/3}/(4m_Q)^{4/3})$ depending on the typical IR mass scale when $m_q$ is small  (\ref{finalrate2})\footnote{Remember that $T$ is the string tension and $m_0$ is the glueball mass scale so in the QCD limit $T/m_0\sim \Lambda_{QCD}$.}.
In the opposite regime where the flux tube is heavy, $M\sim LT$, the relativistic factor (\ref{relativ1}) introduces the relevant dependence $\sqrt{m_Q/M}$, that suppresses the rate.

The decay is ``asymmetric'', that is one of the two decay products has much larger spin than the other.

Finally, the rate is power-like suppressed by the mass of the lighter quarks $m_q$; this power behavior makes this channel the leading one for the decay, at least in the range of parameters specified above.

Even if the theory under consideration is of course not real QCD, we can ask whether at least some of these properties could be expected to be qualitatively the same, with the aim of understanding how far is the string description from the actual strong interactions. 
The first three properties, namely the dependence of the rate on the spin $J$ and on the mass $M$ of the meson, and the asymmetry of the decay, are due to the almost rectangular shape of the string describing the meson, see figure 1.
In a model dual to real QCD this shape could be essentially the same, since it just depends on the background being confining. 
Thus, the behavior of the rate should be qualitatively close to that in QCD (in the planar limit and at strong coupling).
The other possibility is that these features, and in particular the asymmetry of the decay, could be genuine strong coupling effects.
Unfortunately, a comparison with phenomenology is prevented by the lack of experimental data.
In fact, the rate calculated should be suitable to describe the decay of $b \bar b$ mesons into $B$ mesons\footnote{See \cite{evans} for a recent estimation of the B spectrum in another holographic model of strong interactions.} or $c \bar c$ mesons into $D$ mesons.
There are too few experimental results reported on such decays to make a comparison.    
Also, no check of the dependence on the masses of the quarks or the meson can be made, for the same reason.

The last property of the rate, namely the exact dependence on the mass $m_q$ of the lighter quarks, depends heavily on the details of the dual model, so no comparison to QCD can be performed.\\ 

The decay rate for the high spin, small quark mass mesons (\ref{finalrate3}) is linear with the mass $M$ of the meson (so linear with $\sqrt{J}$, since in this regime there are linear Regge trajectories).
This result is equal to the one from the old string picture of hadrons, the only difference being in the coefficients.

The linear dependence on $M$ is expected to go through a real QCD dual on average.
Also, it is well known that 2d QCD exhibits asymptotically this linear behavior on average.
Unfortunately, it is also well known that experimental data are again not conclusive in checking this prediction.

The pre-factor in formula (\ref{finalrate3}) is model dependent and is not expected to be the same in a real QCD dual.\\ 

The picture that emerges of the field theory process in this regime 
is the following.
The flux tube in the mesons made by massive quarks has almost constant energy density everywhere, apart from a small region around the quarks (this just comes from the shape of the string). 
In the decay to massive quarks, the flux tube has enough energy density for pair production only around the quarks, otherwise the energy density is too small (the string goes all the way below the flavor brane): it is a system of a long tube with large energy (because of its length) but small energy density, with highly energetic end-points.
The decay can happen only at the ends of the tube, giving rise to a long, massive meson with large spin and a short, lighter one with smaller spin.
This decay is suppressed by the mass of the produced quarks, because the piece of the tube where the decay can happen (i.e. around the heavy quark) is very small \cite{noi}.
In this note it is shown that the suppression is power-like, as opposed to the exponential suppression of the decay channels involving instantonic transitions, so in this regime the process described here is the leading decay channel.

The flux tube in the mesons made by light quarks has constant energy density everywhere. 
The decay happens by pair-production of light quarks, so every piece of the flux tube has enough energy for the process.
As a result, the 
rate is proportional to the length of the flux-tube, so to the mass of the meson (that is just given by the energy stored in the tube).

\begin{center}    
{\large  {\bf Acknowledgments}}
\end{center}
We thank S. Afonin, M. Cacciari, A. Paredes, A. Pineda, J. Russo, J. Soto, L. Tagliacozzo, P. Talavera, W. Troost and especially L. Martucci for useful discussions and suggestions.
This work is partially  supported by the European
Commission contracts MRTN-CT-2004-005104, MRTN-CT-2004-503369,
CYT FPA 2004-04582-C02-01, CIRIT GC 2001SGR-00065, MEIF-CT-2006-024173.


\end{document}